\documentclass[cameraready]{Interspeech}
\usepackage{amsmath, graphicx, booktabs, multirow, subcaption}

\title{Grammar-Guided Hierarchical Parsing for Long-form Audio Activity Recognition}

\author{Peng}{Zhang}
\author{Qingyu}{Luo}
\author{Philip J.B.}{Jackson}
\author{Wenwu}{Wang}

\address{
    Centre for Vision, Speech and Signal Processing (CVSSP), University of Surrey, U.K.
}

\email{\{p.zhang, qingyu.luo, p.jackson, w.wang\}@surrey.ac.uk}


\usepackage{comment}

\begin{document}

\maketitle

\begin{abstract}
    Long-form audio exhibits an inherent hierarchy: fine-grained events form sub-activities, which in turn constitute higher-level activities. Prior work often models these levels separately, leading to cross-level inconsistencies and requiring supervision at multiple levels. We formulate the problem as hierarchical parsing from event-level evidence: given detected event segments with class posteriors, we infer an order-consistent Act-Sub-Event parse tree. We propose Hierarchical Activity Grammar, encoding hierarchical composition and temporal-order constraints, and perform grammar-guided decoding that combines event evidence with a grammar prior. This yields a temporally grounded parse tree from which sub-activity segmentation and activity classification are derived, without requiring sub-activity or activity labels for training. Experiments on the long-form MultiAct audio dataset demonstrate improved temporal-order consistency (Edit score) and produces interpretable hierarchies.
\end{abstract}

\section{Introduction}
Long-form audio recordings naturally arise in everyday activities such as cooking, cleaning, and tool-based tasks~\cite{huh2025epic,damen2018scaling, zhang2025multiact}. Recent egocentric datasets further underscore the prevalence of such recordings in real-world settings~\cite{song2023ego4d,grauman2022ego4d}. Crucially, these recordings are not merely a flat set of isolated sound events; they are the acoustic trace of a goal-directed process unfolding over time. This process is governed by latent structure, including step boundaries, ordering constraints, and recurring patterns, which organises local acoustic evidence into a coherent sequence. Recovering such structure is essential for building audio systems that move beyond fragmented event-level predictions towards coherent long-form interpretations.

Despite rapid progress in audio recognition, much of the literature has focused on short-term, pre-segmented recordings in everyday environments, spanning clip-level tagging~\cite{gemmeke2017audio, kong2020panns, fonseca2021fsd50k}, acoustic scene classification~\cite{hou2023cooperative, turpault2019sound}, and short-horizon sound event detection~\cite{hu2025pseldnets}. While these formulations are highly effective for learning local acoustic cues and achieving strong event-level recognition, they typically place less emphasis on the long-range temporal structure and hierarchical organisation required for recognising activities from long-form recordings.

However, moving from recognising isolated events to modelling long-form activities introduces a distinct set of challenges. Long-form recordings often span minutes to hours and feature background clutter and diverse acoustic realisations of the same sub-activity. Moreover, event detections may be missing or temporally imprecise, and long-range dependencies across steps are difficult to learn reliably from acoustic observations alone~\cite{adavanne2018sound, wang2019connectionist, zaman2024transformers}. These challenges motivate structured inference that can impose compositional constraints while remaining robust to imperfect event-level evidence, complementing learning-based event detection~\cite{xu2018large, kong2019sound, kong2020sound}.

A recent step towards long-form audio activity recognition is MultiAct~\cite{zhang2025multiact}, which models activities, sub-activities, and events hierarchically. Its design leverages multi-level supervision to learn representations that capture both fine-grained acoustic events and longer-range temporal context, yielding strong per-level baselines on long recordings. However, the three levels are largely optimised with separate objectives and decoded independently, with limited explicit structure tying them together at inference time. In particular, MultiAct does not explicitly impose compositional or ordering constraints across levels. As a result, predictions can be locally plausible yet globally incoherent, for example, when event hypotheses drift across sub-activity boundaries, when sub-activity sequences violate plausible step orderings, or when sub-activity parses are incompatible with the inferred activity.

Beyond purely neural sequence models, prior work in speech and language processing has explored induced symbolic structures for modelling continuous signals and domain-specific semantics. This includes unsupervised acoustic or segmental unit discovery~\cite{bacchiani1995unsupervised,siu2014unsupervised,chaudhuri2012unsupervised}, as well as semi-automatic grammar induction for spoken dialogue understanding~\cite{siu1999semi,wong2001learning,siu2001semi, meng2002semiautomatic,siu2003example}. Motivated by this broader line of structured modelling, recent activity understanding methods incorporate explicit activity grammars and parsing-based inference to enforce hierarchical and temporal constraints~\cite{gong2023activity,ding2023temporal}. Such grammar-guided decoding has been shown to improve interpretability and temporal consistency on long, untrimmed RGB videos by restricting predictions to feasible step orderings and reducing implausible or over-segmented outputs~\cite{yihui2024improving,zhang2025leveraging}. These results motivate our work on grammar-based structured inference for long-form audio activity recognition.

\begin{figure*}[t]
\vspace{-10pt}
  \centering
  \includegraphics[width=\textwidth]{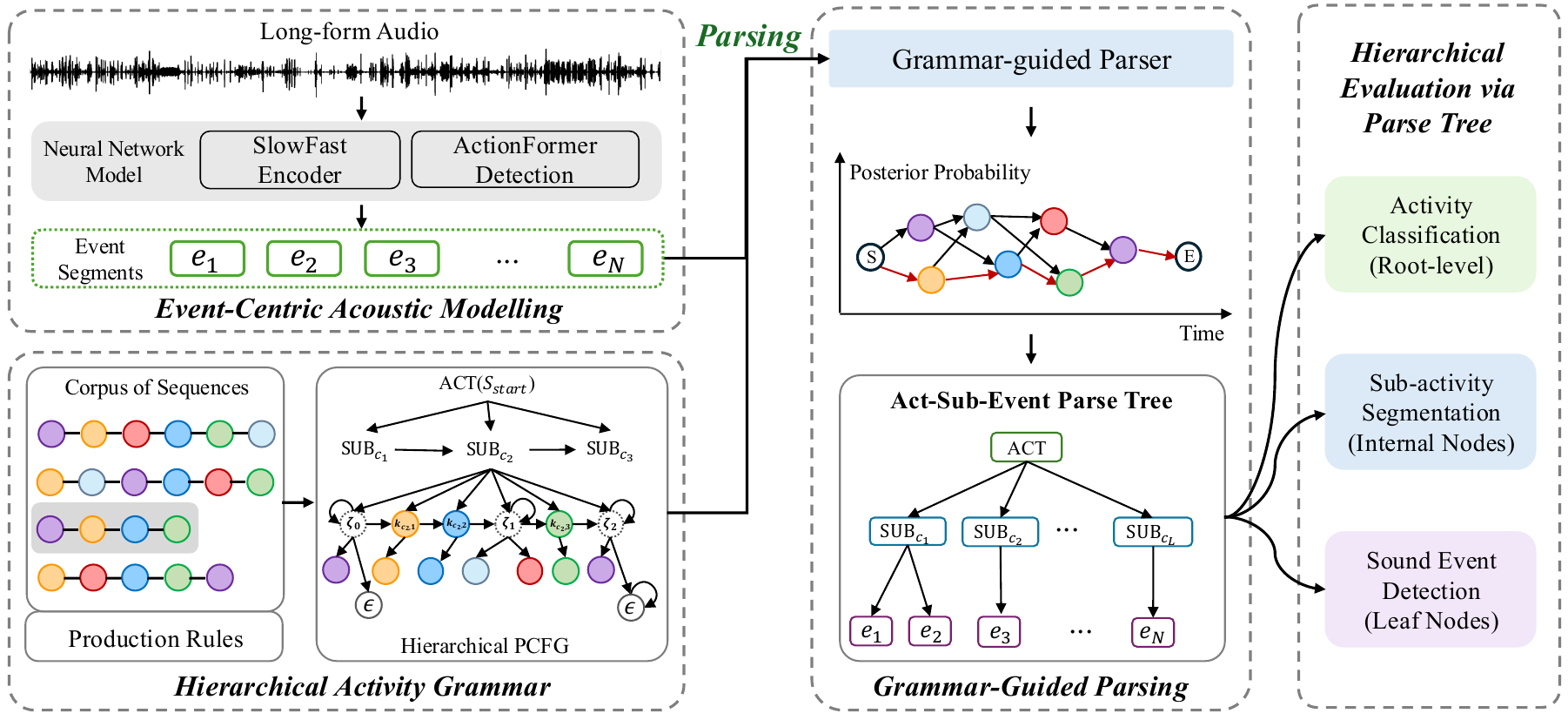}
  \caption{\textbf{Overview of the proposed framework.} Long-form audio is mapped to an ordered event sequence $\{e_i\}_{i=1}^{N}$, which is parsed under our proposed Hierarchical Activity Grammar (HAG) to induce an Act--Sub--Event Parse Tree for multi-level evaluation (Activity/Sub-activity/Event).}
\label{fig:framework}
\vspace{-15pt}
\end{figure*}

As illustrated in Figure~\ref{fig:framework}, we propose a grammar-guided framework for long-form audio activity recognition that performs hierarchical inference from event-level evidence. We formulate the task as hierarchical parsing, which enables sub-activity segmentation and activity classification without relying on sub-activity or activity annotations during training. We induce a Hierarchical Activity Grammar (HAG) from training scripts as a probabilistic context-free grammar (PCFG) that captures hierarchical composition and temporal ordering. The grammar is further augmented with noise non-terminals to absorb missing or spurious events. Finally, we introduce an Earley-style maximum a posteriori (MAP) decoding algorithm for grammar-guided inference, yielding improved temporal-order consistency and competitive hierarchical predictions on the long-form MultiAct audio dataset.

\vspace{-4pt}
\section{Methodology}
\subsection{Problem Definition}
\label{subsec:problem}
\vspace{-4pt}

Given a long-form audio recording $x$, an event-centric acoustic front-end produces an onset-ordered sequence of $N$ detected event segments
\begin{equation}
\setlength{\abovedisplayskip}{4pt}
\setlength{\belowdisplayskip}{4pt}
E=\big((t_n^{s}, t_n^{e}, \pi_n)\big)_{n=1}^{N},
\end{equation}
where $(t_n^{s}, t_n^{e})$ are the onset/offset times of the events and $\pi_n\in[0,1]^{|\Sigma|}$ denotes class posteriors over event labels $\Sigma$.

Our goal is to infer a temporally grounded Act--Sub--Event parse tree $T$: leaves correspond to the event segment indices in $E$, internal nodes correspond to sub-activities spanning contiguous segment ranges, and the root corresponds to the high-level activity. 

Let $\mathcal{T}(E)$ be the set of parse trees that are syntactically valid under a Hierarchical Activity Grammar $\mathcal{G}$ (Sec.~\ref{subsec:hag}) and span the segment sequence $E$. We decode the most plausible tree by combining event-level acoustic evidence with a grammar prior:
\begin{equation}
\setlength{\abovedisplayskip}{4pt}
\setlength{\belowdisplayskip}{4pt}
T^* = \arg\max_{T \in \mathcal{T}(E)}
\Big[ \log p_{\text{acous}}(T \mid x) + \lambda \log p_{\text{gram}}(T) \Big],
\label{eq:map_log}
\end{equation}
where $p_{\text{acous}}$ aggregates segment-level posteriors along the terminals induced by $T$, $p_{\text{gram}}$ factorises over grammar rule probabilities, and $\lambda$ controls the strength of the grammar prior.

We train the event-centric acoustic front-end using event labels and boundaries only. Sub-activity and activity labels are not used for training and are derived from the inferred tree $T^{*}$. We treat event-level predictions as fixed evidence and focus on grammar-guided hierarchical inference.

\vspace{-4pt}
\subsection{Hierarchical Activity Grammar}
\vspace{-4pt}
\label{subsec:hag}

Inspired by the hierarchical structure of language~\cite{chomsky1956three}, we model long-form  audio as a structured composition of semantic units, in which a complex activity is decomposed into a sequence of sub-activities and their constituent sound events. To capture such multi-level dependencies and the underlying procedural syntax, we propose the HAG to model the structural organisation of long-form audio activities across multiple time scales.

We formulate HAG as a PCFG, which models the hierarchical decomposition of long-form audio activities into sub-activities and events through terminal and non-terminal nodes. Terminal nodes correspond to observed audio events, while non-terminal nodes represent latent sub-activities/activities and their composition rules, allowing us to score and select plausible event-to-activity decompositions under procedural constraints. Formally, we define the grammar as $\mathcal{G} = (\mathcal{N}, \Sigma, \mathcal{R}, S_{\text{start}}, P)$, where the non-terminal set $\mathcal{N} = \mathcal{A} \cup \mathcal{S} \cup \mathcal{N}_{\text{noise}}$ encodes a latent hierarchy of high-level activities ($\mathcal{A}$), intermediate sub-activities ($\mathcal{S}$), and transient noise nodes ($\mathcal{N}_{\text{noise}}$) for structural robustness. Terminals $\Sigma$ correspond to sound event classes predicted by the event-centric acoustic model. The start symbol $S_{\text{start}} \in \mathcal{A}$ denotes the root activity for a given recording. $P$ denotes the production probabilities, normalised for each left-hand-side non-terminal. Concretely, for any non-terminal $X$, we normalise over all productions $X\rightarrow \alpha$ (where $\alpha$ is a sequence of terminals/non-terminals) such that $\sum_{\alpha:(X\rightarrow\alpha)\in\mathcal{R}} P(X\rightarrow\alpha)=1$.

To bridge high-level logic and acoustic observations, we decompose the production rule set into two layers, $\mathcal{R}=\mathcal{R}_{\text{act}} \cup \mathcal{R}_{\text{sub}}$, where $\mathcal{R}_{\text{act}}$ captures global temporal logic and $\mathcal{R}_{\text{sub}}$ models local acoustic variability. Let $\mathcal{S}=\{\text{SUB}_c\}_{c=1}^{|\mathcal{S}|}$ denote the set of sub-activity non-terminals, indexed by sub-activity type $c$. The first layer expands the root activity $S_{\text{start}}$ (denoted as $\text{ACT}$) into an ordered sequence of $L$ sub-activities:
\begin{equation}
\setlength{\abovedisplayskip}{2pt}
\setlength{\belowdisplayskip}{4pt}
\text{ACT} \rightarrow \text{SUB}_{c_1}\cdots \text{SUB}_{c_L}, \quad c_\ell \in \{1,\ldots,|\mathcal{S}|\}.
\end{equation}
Here, $\ell \in \{1,\ldots,L\}$ indexes the position in the sub-activity sequence, and $c_\ell$ specifies the sub-activity type assigned to the $\ell$-th position. The type sequence $(c_1,\dots,c_L)$ has length $L$ and may contained repeated types.

At the sub-activity level, $\mathcal{R}_{\text{sub}}$ captures local acoustic variability by anchoring each sub-activity non-terminal $\text{SUB}_c\in\mathcal{S}$ to a small set of characteristic event classes. 
We denote the anchor set by $K_c\subseteq\Sigma$, with $|K_c|=M_c$, which provides an expected acoustic backbone for $\text{SUB}_c$. All remaining event classes are treated as optional non-anchor events, since they often arise from incidental/background sounds or spurious detections, and we denote this set by $U_c := \Sigma \setminus K_c$. To allow such optional events without disrupting the anchor structure, we introduce noise non-terminals $\zeta_i\in\mathcal{N}_{\text{noise}}$ for $i=0,\ldots,M_c$, which can appear before, between, and after anchors:
\begin{equation}
\setlength{\abovedisplayskip}{2pt}
\setlength{\belowdisplayskip}{2pt}
\text{SUB}_c  \rightarrow \zeta_0\, k_{c,1}\, \zeta_1\, k_{c,2}\, \cdots\, k_{c,M_c}\, \zeta_{M_c}.
\end{equation}

Each noise node $\zeta_i$ generates a (possibly empty) sequence over $U_c$ via
\begin{equation}
\setlength{\abovedisplayskip}{2pt}
\setlength{\belowdisplayskip}{2pt}
\zeta_i \rightarrow u\,\zeta_i \ \mid\ \epsilon,\qquad u \in U_c,
\end{equation}
where at each expansion $u$ can be any terminal in $U_c$, and $\epsilon$ denotes the empty string.

\begin{figure}[t]
  \centering
  \includegraphics[width=1.0\linewidth]{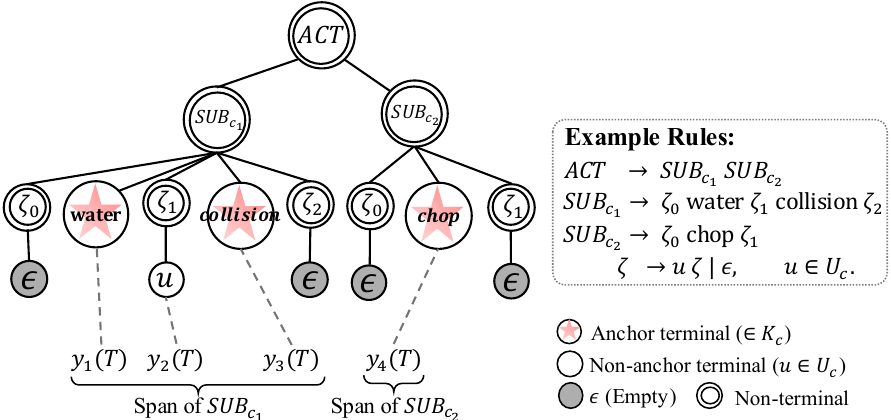}
  \caption{Running Example of HAG.}
  \label{fig:hag-example}
  \vspace{-20pt}
\end{figure}

To clarify the above formulation, Figure~\ref{fig:hag-example} illustrates a concrete parse produced by our grammar and maps each symbol to a visual element. The root non-terminal $\text{ACT}$ expands into an ordered sequence of sub-activities ($\text{SUB}_{c_1}, \text{SUB}_{c_2}$) according to $\mathcal{R}_{\text{act}}$, while each $\text{SUB}_c$ further expands under $\mathcal{R}_{\text{sub}}$ into anchor terminals (e.g., \textit{water}, \textit{collision}, \textit{chop}) interleaved with noise non-terminals $\zeta_i$ that optionally emit non-anchor terminals $u$ or $\epsilon$. The resulting parse tree also induces contiguous temporal spans for each sub-activity, as indicated at the bottom of the figure.

Finally, we equip the grammar with production probabilities. Given a parse tree $T$, its grammar probability factorises over the productions applied in the derivation:
\begin{align}
\setlength{\abovedisplayskip}{2pt}
\setlength{\belowdisplayskip}{2pt}
p_{\text{gram}}(T) &= \prod_{r\in T} P(r) \\
&= \Big(\prod_{r\in T\cap \mathcal{R}_{\text{act}}} P(r)\Big)
   \Big(\prod_{r\in T\cap \mathcal{R}_{\text{sub}}} P(r)\Big).
\end{align}
By combining $p_{\text{gram}}(T)$ as a structural prior with acoustic evidence during grammar-guided parsing, the framework resolves ambiguities in the detected event sequence and yields a coherent hierarchical representation without level-specific neural modelling.

\vspace{-4pt}
\subsection{Event-Centric Acoustic Modelling}
\vspace{-4pt}
\label{subsec:acoustic}

We adopt an event-centric formulation where discrete sound events are the fundamental units. Let
\begin{equation}
\setlength{\abovedisplayskip}{4pt}
\setlength{\belowdisplayskip}{4pt}
Y(T)=(y_1(T),\ldots,y_N(T))
\end{equation}
denote the terminal label sequence induced by a parse tree $T$ over the ordered segment sequence $E$, where each leaf aligns with one detected segment and $y_n(T)\in\Sigma$ is its event class. In Figure~\ref{fig:hag-example}, $Y(T)=(\texttt{water},\,u,\,\texttt{collision},\,\texttt{chop})$, with $u\in U_c$ generated by a noise node $\zeta_i$.

For each segment $n$, the event detector outputs a posterior vector $\pi_n\in[0,1]^{|\Sigma|}$ over terminal classes in $\Sigma$. We score a tree $T$ by aggregating posteriors along its terminals:
\begin{equation}
\setlength{\abovedisplayskip}{2pt}
\setlength{\belowdisplayskip}{2pt}
p_{\text{acous}}(T\mid x)\ \triangleq\ \prod_{n=1}^{N} \pi_n\big(y_n(T)\big).
\label{eq:p_acous}
\end{equation}
This corresponds to a conditional-independence assumption across segments conditioned on the posteriors produced by our event detector. For efficiency, during scanning we restrict candidate terminals to the top-$m$ classes per segment, without changing the decoding objective.

In practice, we adopt the event-level detector from the MultiAct baseline\footnote{\url{https://github.com/PennyZhang9/MultiAct}}, implemented with a SlowFast-based~\cite{kazakos2021slow} encoder and an ActionFormer-based~\cite{zhang2022actionformer} segmenter, to obtain event segments and their class posteriors $\pi_n$. 
Different from MultiAct's hierarchical design~\cite{zhang2025multiact} that further predicts sub-activities and activities with level-specific heads, our proposed framework restricts neural modelling to the event level; higher-level organisation is recovered by grammar-guided parsing under HAG.

\vspace{-4pt}
\subsection{Grammar-Guided Parsing}
\vspace{-4pt}
\label{subsec: parser}
To maximise Eq.~\eqref{eq:map_log}, we adopt a Viterbi-style Earley parser~\cite{stolcke1995efficient, qi2018generalized} adapted to HAG. We perform decoding in the log domain, maintaining for each parser state the maximum partial score and a backpointer for reconstructing the optimal parse tree $T^*$. The parser iterates three operations with corresponding log-score updates:
\begin{itemize}
    \item \textbf{Predict:} expand a non-terminal by applying a production $r$ and add $\lambda \log P(r)$.
    \item \textbf{Scan:} match a terminal event label $y \in \Sigma$ at segment position $n$ and add $\log \pi_n(y)$; for efficiency, we prune candidates to the top-$m$ terminal labels per segment with $m{=}10$ (Sec.~\ref{subsec:acoustic}).
    \item \textbf{Complete:} once a non-terminal has been fully matched, use it to update any partial parses that expected it and keep only the highest-scoring hypothesis for each resulting parser state via a Viterbi max update.
\end{itemize}

The best completed derivation spanning all $N$ segments yields $T^*$ via back track. We then extract spans for completed $\text{ACT}$ and $\text{SUB}_c$ constituents and map a segment range $[i,j]$ to real time as $[t_i^{s},\, t_j^{e}]$; the predicted activity label is the root $\text{ACT}$ in $T^*$.

\vspace{-4pt}
\section{Experiments}

\vspace{-4pt}
\subsection{Dataset}
\vspace{-4pt}

Experiments are conducted on MultiAct~\cite{zhang2025multiact}, a long-form procedural audio dataset whose aligned activity-, sub-activity-, and event-level annotations directly match our Act--Sub--Event parsing formulation. Unlike many larger clip-level or flat event benchmarks, MultiAct provides the intermediate procedural structure needed to evaluate hierarchical parsing, including step boundaries, ordering regularities, and long-range dependencies. The dataset comprises 8.97 hours of audio, covering 3 activity classes (51 instances; mean 628\,s), 12 sub-activity classes (472 instances; mean 63.6\,s), and 44 event classes (7312 instances; mean 4.91\,s). It also exhibits substantial duration variance, with sequences lasting up to 3164\,s, making it a strong testbed for structure-based inference on long-form recordings.

\vspace{-4pt}
\subsection{Evaluation Protocol and Results}
\vspace{-4pt}

We evaluate event detection, sub-activity segmentation, and activity classification under an \emph{event-only} setting, using a shared event detector as the sole acoustic evidence. The HAG structure and production probabilities are induced only from training scripts, fixed for Val/Eval, and used solely as a parsing prior over event hypotheses. No validation/evaluation scripts are used to construct grammar rules, and no sub-activity or activity classifiers are trained. Sub-activities and activities are recovered as latent variables from the highest-scoring parse tree.

We report metrics at three levels. For \textbf{event detection}, we compute Average Precision (AP) at multiple temporal IoU (tIoU) thresholds. For \textbf{sub-activity segmentation}, we report the segmental Edit score, segment-level F1 at IoU thresholds 10/25/50, and frame-wise accuracy. Following~\cite{ding2023temporal}, Edit is defined as a normalised Levenshtein distance between predicted and ground-truth segment label sequences $X$ and $Y$: $\mathrm{Edit}=100\cdot\left(1-\frac{e(X,Y)}{\max(|X|,|Y|)}\right)$. For \textbf{activity classification}, we report Top-1 accuracy and macro-averaged metrics across classes, including mPCA (mean per-class accuracy), mAP (mean average precision), and mAUC (mean area under the ROC curve).

\vspace{-5pt}
\subsubsection{Event Detection}
\vspace{-4pt}

As shown in Table~\ref{tab:event}, applying grammar-guided decoding after event detection yields a small but consistent AP improvement over the standalone detector. This is expected because parsing is performed after detection and does not alter event predictions; the slight gains mainly confirm that our decoding introduces no degradation in event-level performance.

\begin{table}[t]
\centering
\footnotesize
\caption{Event detection performance measured by Average Precision (AP) at different temporal IoU thresholds (\%, $\uparrow$).}
\label{tab:event}
\scalebox{0.8}{ 
\begin{tabular}{cccccccc}
\toprule
\multirow{2}{*}{Split} & \multirow{2}{*}{Method} &
\multicolumn{5}{c}{AP @ tIoU} & \multirow{2}{*}{mAP} \\
\cmidrule(lr){3-7}
 &  & 0.1 & 0.2 & 0.3 & 0.4 & 0.5 &  \\
\midrule
\multirow{2}{*}{Val}
 & Event NN$^{\dagger}$                 & 16.98 & 14.70 & 12.81 & 11.18 &  9.84 & 13.10 \\
 & Event NN + Grammar$^{\ddagger}$       & \textbf{17.00} & \textbf{14.79} & \textbf{12.83} & \textbf{11.19} &  \textbf{9.85} & \textbf{13.13} \\
\midrule
\multirow{2}{*}{Eval}
 & Event NN$^{\dagger}$                 & 16.48 & 15.66 & 14.97 & 13.55 & 12.49 & 14.63 \\
 & Event NN + Grammar$^{\ddagger}$       & \textbf{16.50} & \textbf{15.68} & \textbf{14.99} & \textbf{13.57} & \textbf{12.50} & \textbf{14.65} \\
\bottomrule
\end{tabular}
}
{\scriptsize $^{\dagger}$ Reported in MultiAct~\cite{zhang2025multiact}. $^{\ddagger}$ Evaluated by us with grammar-guided decoding applied after detection.}
\vspace{-10pt}
\end{table}

\vspace{-5pt}
\subsubsection{Sub-activity Segmentation}
\vspace{-4pt}

\begin{table}[t]
\centering
\footnotesize
\caption{Sub-activity segmentation results (\%, $\uparrow$). \textbf{Fully-supervised} method uses sub-activity boundary supervision and serves as an upper bound. \textbf{Event-only} methods use only event predictions as evidence (no sub-activity labels for training). }
\label{tab:subact}
\scalebox{0.75}{ 
\begin{tabular}{llcccccc}
\toprule
Split & Method & Sup. & Edit & F1@10 & F1@25 & F1@50 & Acc. \\
\midrule
\multicolumn{8}{l}{\textit{Fully-supervised baseline}}\\
Val  & Subactivity NN$^{\ddagger}$ & E+S & 36.3 & 39.5 & 33.5 & 18.0 & 38.9 \\
Eval & Subactivity NN$^{\ddagger}$ & E+S & 24.6 & 21.9 & 17.4 & 9.0  & 21.3 \\
\midrule
\multicolumn{8}{l}{\textit{Event-only inference}}\\
Val  & Grammar-induced (ours) & E & \textbf{37.1}$\blacktriangle$ & 24.6 & 21.1 & 15.8 & 25.5 \\
Eval & Grammar-induced (ours) & E & \textbf{35.3}$\blacktriangle$ & \textbf{24.8}$\blacktriangle$ & 14.0 & 7.8  & 19.7 \\
\bottomrule
\end{tabular}
}
{\scriptsize Sup.: E = event-only; E+S = event + sub-activity boundary supervision. $^{\ddagger}$ Evaluated by us using the official MultiAct code/protocol.}
\vspace{-16pt}
\end{table}

Table~\ref{tab:subact} shows a clear trade-off between structural consistency and boundary-level localisation. Our grammar-induced decoding yields consistent gains in Edit on both splits, with a particularly large improvement on Eval (24.6 $\rightarrow$ 35.3), indicating that HAG effectively regularises the global sub-activity sequence. In contrast, overlap-based metrics at stricter IoU thresholds (F1@25/F1@50) and frame-wise accuracy do not improve, and in several cases degrade. This behaviour is expected because our parser operates on the event detector's segment sequence: it can reassign labels and merge/suppress spurious transitions via grammatical constraints, but it cannot refine temporal boundaries beyond the detector's proposal granularity. As a result, when the correct sub-activity order is recovered but boundary alignment remains coarse, Edit increases while high-IoU F1 remains limited.

Overall, the results highlight HAG’s effectiveness in enforcing global procedural structure, and suggest that more accurate event proposals and better boundary alignment are key to further improving overlap-based segmentation performance.

\vspace{-4pt}
\subsubsection{Activity Classification}
\vspace{-4pt}

\begin{table}[t]
\centering
\footnotesize
\caption{High-level activity classification results (\%, $\uparrow$). \textbf{Fully-supervised} method uses sub-activity labels and activity labels, serves as a fully-supervised upper bound. \textbf{Event-only} grammar induced: event + grammar $\rightarrow$ sub-activity $\rightarrow$ root activity.}
\label{tab:activity}
\scalebox{0.80}{
\begin{tabular}{llccccc}
\toprule
Split & Method & Sup. & Top-1 & mPCA & mAP & mAUC \\
\midrule
\multicolumn{7}{l}{\textit{Fully-supervised baseline}}\\
Val  & Activity NN$^{\dagger}$ & E+S+A & 66.7 & 61.9 & 72.7 & 84.6 \\
Eval & Activity NN$^{\dagger}$ & E+S+A & 83.3 & 83.3 & 72.2 & 70.8 \\
\midrule
\multicolumn{7}{l}{\textit{Event-only hierarchical inference}}\\
Val  & Grammar-induced (ours) & E & \textbf{73.3}$\blacktriangle$ & \textbf{70.3}$\blacktriangle$ & 59.5 & 78.5 \\
Eval & Grammar-induced (ours) & E & 66.7 & 66.7 & 58.3 & \textbf{75.0}$\blacktriangle$ \\
\bottomrule
\end{tabular}
}
{\scriptsize Sup.: E = event-only; E+S+A = event + sub-activity + activity label supervision. $^{\dagger}$ Reported in MultiAct~\cite{zhang2025multiact}.}
\end{table}

We derive the activity label as the root of the highest-scoring parse, performing event-only hierarchical inference (event $\rightarrow$ sub-activity $\rightarrow$ activity). As shown in Table~\ref{tab:activity}, grammar-induced inference is strong on Val (73.3\% Top-1, 70.3 mPCA) but drops on Eval. Despite not surpassing the fully supervised method on all metrics, our approach still achieves 66.7\% Top-1 accuracy and 75.0\% mAUC on Eval without using any sub-activity or activity labels, demonstrating competitive root-level inference from structured decoding. Improving robustness to imperfect evidence remains a direction for future work.

\begin{figure}[t]
  \centering
  \vspace{-7pt}
  \begin{subfigure}[b]{0.48\linewidth}
    \centering
    \includegraphics[width=\linewidth]{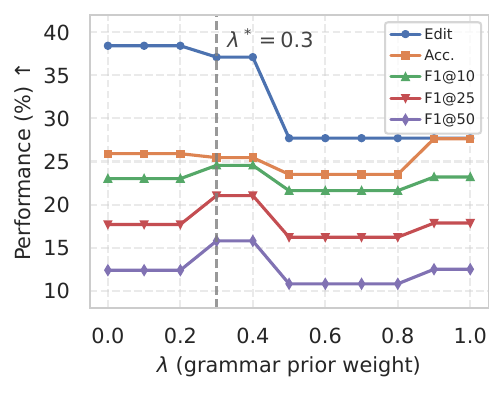}
    \caption{Grammar weight $\lambda$ sweep.}
    \label{fig:lambda_sweep}
  \end{subfigure}
  \hfill
  \begin{subfigure}[b]{0.48\linewidth}
    \centering
    \raisebox{5pt}{\includegraphics[width=\linewidth]{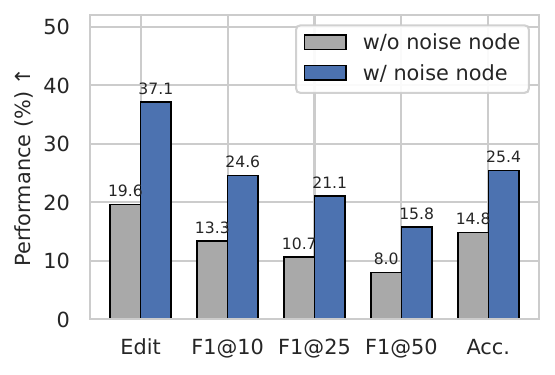}}
    \caption{Impact of noise node.}
    \label{fig:ablation_noise}
  \end{subfigure}
  \caption{Empirical analysis on the validation set.}
  \vspace{-20pt}
\end{figure}

\vspace{-5pt}
\subsubsection{Ablation Studies}
\vspace{-4pt}

\noindent\textbf{Effect of Grammar Weight.}
Fig.~\ref{fig:lambda_sweep} shows that moderate $\lambda$ ($\approx 0.3$--$0.4$) yields consistently strong performance across metrics, while large $\lambda$ degrades results, indicating over-regularisation by the grammar prior. We thus fix $\lambda=0.3$ in all experiments.

\noindent\textbf{Impact of Noise Nodes.}
We ablate HAG noise non-terminals that handle non-anchor events and missed detections. As shown in Fig.~\ref{fig:ablation_noise}, adding noise nodes consistently improves all metrics, indicating they absorb spurious segments and mitigate error propagation in parsing.

\vspace{-7pt}
\section{Conclusion}
\vspace{-4pt}

In this paper, we presented a grammar-guided hierarchical parsing framework for long-form audio activity recognition. By combining a hierarchical activity grammar with event-level evidence, our decoder enforces compositional structure and temporal ordering while yielding interpretable Act--Sub--Event parse trees. On the long-form MultiAct audio dataset, grammar-guided decoding improves temporal-order consistency (Edit score) without training additional sub-activity or activity classifiers, and noise non-terminals further enhance robustness to imperfect event proposals. Remaining challenges include boundary-accurate sub-activity segmentation and sensitivity to the grammar prior weight, motivating more robust evidence integration and broader procedural settings.

\section{Acknowledgments}
We thank Pablo Martínez-Nuevo, Sven Ewan Shepstone and Jon Francombe (Bang \& Olufsen A/S) for valuable discussions. This research was supported by Bang \& Olufsen A/S as part of the AURIC Project.

\section{Generative AI Use Disclosure}
Generative AI tools were used only for language editing and improving the readability of the manuscript. No generative AI tools were used to produce the scientific content, methodology, experiments, results, or conclusions. The authors are fully responsible for the content, claims, methodology, experiments, and conclusions of this paper.

\bibliographystyle{IEEEtran}
\bibliography{mybib}

\end{document}